\documentclass[12pt]{article}
\usepackage{multicol}
\usepackage{epsfig}
\usepackage{graphicx}
\usepackage[all,knot]{xy}

\usepackage{amsmath}
\usepackage{color}
\usepackage{url}
\usepackage{multimedia}
\usepackage{setspace} 
\usepackage{epstopdf}

\newcommand\pubnumber{CIPANP2015-Healey}
\newcommand\pubdate{\today}

\def\torino{Department of Physics\\
Universita di Torino, I-10125 Torino, ITALY\\
INFN Torino, ITALY}

\def\Title#1{\begin{center} {\Large #1 } \end{center}}
\def\Author#1{\begin{center}{ \sc #1} \end{center}}
\def\Address#1{\begin{center}{ \it #1} \end{center}}

\newcommand\pubblock{\rightline{\begin{tabular}{l} \pubnumber\\
         \pubdate  \end{tabular}}}
\newenvironment{Abstract}{\begin{quotation}  }{\end{quotation}}
\newenvironment{Presented}{\begin{quotation} \begin{center} 
             PRESENTED AT\end{center}\bigskip 
      \begin{center}\begin{large}}{\end{large}\end{center} \end{quotation}}

\newcommand{\vcb}{$V_{cb}\ $}
\newcommand{\mvcb}{$|V_{cb}|\ $}
\newcommand{\as}{\alpha_s}
\newcommand{\mupi}{\mu_\pi^2}
\newcommand{\mug}{\mu_G^2}
\newcommand{\rd}{\rho_D^3}
\newcommand{\rls}{\rho_{LS}^3}

\def \be{\begin{equation}}
\def \ee{\end{equation}}
\newcommand{\bea}{\begin{eqnarray}}
\newcommand{\eea}{\end{eqnarray}}
\newcommand{\bem}{\begin{multline}}
\newcommand{\eem}{\end{multline}}

\def\beq{\begin{equation}}
\def\eeq{\end{equation}}
\def\beqa{\begin{eqnarray}}
\def\eeqa{\end{eqnarray}}

\def\beq{\begin{equation}}
\def\eeq{\end{equation}}
\def\beqa{\begin{eqnarray}}
\def\eeqa{\end{eqnarray}}

\definecolor{red}{rgb}{0.5,0,0}
\definecolor{blue}{rgb}{0,0,0.5}
\definecolor{green}{rgb}{0,0.3,0}
 
\definecolor{Red}{rgb}{1,0,0}
\definecolor{Blue}{rgb}{0,0,1}
\definecolor{teal}{rgb}{0,0,1}
\definecolor{Green}{rgb}{0,1,0}
\definecolor{magenta}{rgb}{1,0,.6}
\definecolor{lightblue}{rgb}{0,.5,1}
\definecolor{lightpurple}{rgb}{.6,.4,1}
\definecolor{gold}{rgb}{.6,.5,0}
\definecolor{orange}{rgb}{1,0.4,0}
\definecolor{hotpink}{rgb}{1,0,0.5}
\definecolor{newcolor2}{rgb}{.5,.3,.5}
\definecolor{newcolor}{rgb}{0,.3,1}
\definecolor{newcolor3}{rgb}{1,0,.35}
\definecolor{darkgreen1}{rgb}{0, .35, 0}
\definecolor{darkgreen}{rgb}{0, .6, 0}
\definecolor{darkred}{rgb}{.75,0,0}

\begin{document}
\begin{titlepage}
\pubblock

\vfill
\Title{A Precision Determination of the CKM Element $|V_{cb}|$ and a Review of the Status of Inclusive $|V_{ub}|$}
\vfill
\Author{ Kristopher J. Healey}
\Address{\torino}
\vfill
\begin{Abstract}
In these proceedings for CIPANP2015 we present a brief overview of the current status of the theoretical approaches used by our group for the extraction of $|V_{cb}|$ and $|V_{ub}|$ through inclusive semi-leptonic $B$ decays. We discuss the calculations and implications of the recent perturbative corrections to power-suppressed contributions for  $|V_{cb}|$, and present an overview of the major sources of theoretical uncertainty for $|V_{ub}|$.
\end{Abstract}
\vfill
\begin{Presented}
CIPANP2015\\
Twelfth Conference on the Intersections of Particle and\\
Nuclear Physics\\
Vail, CO, U.S.A., May 19-24, 2015
\end{Presented}
\vfill
\end{titlepage}
\def\thefootnote{\fnsymbol{footnote}}
\setcounter{footnote}{0}

\section{Introduction to the Inclusive $|V_{cb}|$ Determination}

A precise theoretical determination of the CKM matrix element \vcb  is imperative for an accurate exploration of heavy flavor phenomena. The $b \to c$ transition is important for the analysis of CP violation in the Standard Model(SM), and in constraining both flavor violating processes and the CKM unitarity triangle values of $\bar{\rho}$ and $\bar{\eta}$. 

The determination of $|V_{cb}|$ from inclusive semileptonic $B$ decays is based on an Operator Product Expansion (OPE) that allows us to express the  widths and the moments of the kinematic distributions of $B\to X_{u,c} \ell\nu$ as double expansions in $\as$ and $\Lambda_{\rm QCD}/m_b$. These corrections are now known perturbatively to $O(\alpha_s^2)$\cite{melnikov,melnikov2,Pak:2008cp,NNLO,btoc}, $O(\alpha_s \Lambda^2_{\rm QCD}/m_b^2)$\cite{Becher:2007tk,Alberti:2013kxa, Alberti:2012dn}, and to $O(\Lambda_{\rm QCD}^3/m_b^3)$\cite{1mb2,1mb3} in the Heavy Quark Expansion (HQE)\cite{paz}.
\small 
\begin{eqnarray}
\label{eq:expan}
\qquad M_i&=&\textcolor{darkgreen}{\qquad \qquad M_i^{(0)} + M_i^{(\pi,0)} \frac{\mu_\pi^2}{m_b^2}+ M_i^{(G,0)} \frac{\mu_G^2}{m_b^2}
+ M_i^{(D,0)} \frac{\rho_D^3}{m_b^3}+ M_i^{(LS,0)} \frac{\rho_{LS}^3}{m_b^3} }\\ \nonumber
&&+ \left(\frac{\alpha_s}{\pi}\right)\left[ \textcolor{darkgreen}{M_i^{(1)}} \textcolor{teal}{+ \,\,M_i^{(\pi,1)} \frac{\mu_\pi^2}{m_b^2}+ M_i^{(G,1)} \frac{\mu_G^2}{m_b^2}}
\textcolor{red}{+ \,M_i^{(D,1)} \frac{\rho_D^3}{m_b^3}+ M_i^{(LS,1)} \frac{\rho_{LS}^3}{m_b^3}} \right]\\ \nonumber
&&+\left(\frac{\alpha_s}{\pi}\right)^2 \textcolor{darkgreen}{M_i^{(2)}} + \textcolor{red}{\mathcal{O}\left(m_b^{4,5}, \alpha_s^3 \right)}\normalcolor
\end{eqnarray}
\normalsize
In Equation \ref{eq:expan} one can see the different elements of the double expansion for a given observable. For the observables that concern us, the green terms have been calculated previously, the blue terms are the new corrections that have been added to the calculations of the necessary observables, and the red terms are currently being calculated or have been approximated in various methods\cite{Mannel:2010wj}. 
Each observable is dependent on the masses of the heavy quarks, $m_b$ and  $m_c$, $\alpha_s$ the strong coupling constant, and the matrix elements of local operators operating on the $B$-meson at increasing powers of $1/m_b$. To the order currently calculated this includes $\mupi$ and $\mug$ at $O(1/m_b^2)$, $\rd$  and  $\rls$ at $O(1/m_b^3)$. These matrix elements can be constrained by various measurements of the first 3 central moments of the lepton energy and hadronic mass distributions of $B\to X_c \ell \nu$. Each observable is measured varying the low-energy cut-off of the leptonic spectrum and these have been measured to good accuracy at the $B$-factories, CLEO, DELPHI and CDF.

Combining with the total semileptonic width, these parameters can then be used to extract \mvcb. In the past this strategy has been quite successful and has allowed for a $\sim 2\%$ determination of \vcb from inclusive decays \cite{HFAG}. 
Additional motivation for increasing the precision of the extracted parameters is a desire to resolve a $\sim2\sigma$ discrepancy that exists between the current inclusive determination and the most precise \mvcb determination from the exclusive $B\to D^* \ell\nu$ at zero recoil with a lattice calculation of the form-factor \cite{HFAG,lattice}. It should be noted however that the zero-recoil form-factor estimate based on heavy quark sum rules leads to $|V_{cb}|$ in good agreement with the inclusive result \cite{long}.  

\subsection{Theoretical Error and Correlations}
The previous procedures used in the semileptonic fits\cite{babarfit,1S,BF} have been recently re-examined and a few relevant issues are worthy of ones concern :  $i)$ the theoretical uncertainties and how they are implemented in the fit, and $ii)$ the inclusion of additional constraints on the parameters. For a thorough overview we direct the reader to consult \cite{Gambino:2013rza}. In brief, considering 100\% correlation between an observable and the same observable with a different cut on the leptonic energy is too strong of a stipulation. This requirement is relaxed and tested with various parametrizations of the correlations between observables. 

We use a limited selection of the theoretical correlation options and additional mass constraints. The full results have been published earlier this year \cite{Alberti:2014yda}. We choose to look solely at the scenario that provided the most accurate and reliable results at $\mathcal{O}(\alpha_s^2)$, which is the case of a functional theoretical correlation between moments \textbf{[D]}, and using the constraints of $m_c = \bar{m}_c(3 GeV)$, and not including external $m_b$ constraints. This is performed with $\alpha_s(\mu) = \alpha_s(m_b) = 0.22$, and $\mu_G^2(\mu_{\mu_G}) = \mu_G^2(m_b)$. 
 
\subsection{Higher Order Corrections}

The accuracy and reliability of the inclusive method depends our ability to control higher order corrections. The perturbative \textcolor{darkgreen}{$\mathcal{O}(\alpha_s^n)$} corrections are known completely to NNLO \cite{melnikov,melnikov2,Pak:2008cp,NNLO}. The \textcolor{red}{$\mathcal{O}(1/m_b^{4,5})$}  higher order non-perturbative corrections have also been calculated\cite{Mannel:2010wj}. The \textcolor{teal}{$\mathcal{O}(\alpha_s/m_b^2)$} cross-terms from the dual expansions are complete \cite{Becher:2007tk,Alberti:2013kxa,Alberti:2012dn}, and implemented in the current fit. The \textcolor{red}{$\mathcal{O}(\alpha_s/m_b^3)$} corrections are currently being calculated.

\subsubsection{$\mathcal{O}(1/m_b^{4,5})$}
  
The \textcolor{red}{$\mathcal{O}(1/m_b^{4,5})$} corrections unfortunately have too many parameters to include in the global fit. 
\begin{eqnarray}
\nonumber 2 M_B m_1 &= \langle ((\vec{p})^2)^2 \rangle \qquad \qquad  &(...) \\
2 M_B m_2 &= \langle g^2\vec{E}^2 \rangle \qquad \qquad  &2 M_B m_8 = \langle (\vec{S} \cdot \vec{B})(\vec{p})^2) \rangle \\
\nonumber  2 M_B m_3 &= \langle g^2\vec{B}^2 \rangle \qquad \qquad  &2 M_B m_9 = \langle \Delta (\vec{\sigma} \cdot \vec{B}) \rangle
 \end{eqnarray}

Their contribution has been estimated using the Lowest Lying State Saturation \cite{Mannel:2010wj,Heinonen:2014dxa}. This is done by truncating the series of insertions of the lowest lying states,
\beq
\langle B | \mathcal{O}_1 \mathcal{O}_2 | B\rangle = \sum_n\langle B | \mathcal{O}_1  | n\rangle \langle n | \mathcal{O}_2 | B\rangle.
\eeq

The influence of these corrections on $|V_{cb}|$ have been explored by S. Turczyk and P. Gambino and preliminary results are around $\frac{\delta V_{cb}}{V_{cb}}\simeq-0.35\%$. LLSA corrections have also been found, but allowing $80\%$ gaussian deviations from LLSA seem to leave $V_{cb}$ unaffected.

\subsubsection{$\mathcal{O}(\alpha_s/m_b^2)$}
 
The newest corrections to be included are the NLO perturbative corrections to the kinetic operators at $1/m_b^2$. The goal of the calculation is to obtain the contributions to the hadronic tensor,

\begin{equation}
W^{\mu \nu} = \frac{(2 \pi)^3}{2 m_B} \sum_{X_c} \delta^4(p_b - q- p_X)\langle B|J_L^{\dagger \mu} X_c\rangle\langle X_c | J_L^{\nu} | B \rangle.
\end{equation}
This can be decomposed into
\begin{equation}
\nonumber m_b W^{\mu \nu} = -W_1 g^{\mu \nu}  + W_2 v^{\mu} v^{\nu} + i W_3 \epsilon^{\mu \nu \rho \sigma} v_{\rho} \hat{q}_{\sigma} + (...),
\end{equation}
where each $W_i$ can be double-expanded in $1/m_b$ and $\alpha_s$,
\begin{equation}
\nonumber W_i = W_i^{(0)} + \frac{\mu^2_{\pi,G}}{2 m_b^2} W_i^{(\pi,G,0)}+ \frac{C_F \alpha_s}{\pi}   \left[W_i^{(1)} + \textcolor{teal}{\frac{\mu^2_{\pi,G}}{2 m_b^2} W_i^{(\pi,G,1)} }\right] .
\end{equation}

A useful check are the re-parametrization relations for $W_i^{\pi,n}$, where for all orders

\beq
W_3^{(\pi,n)} = \frac{5}{3} \hat{q}_0 \frac{dW_3^{(n)}}{d\hat{q}_0} - \frac{\hat{q}^2-\hat{q}_0^2}{3} \frac{dW_3^{(n)}}{d\hat{q}_0}.
\eeq

\begin{figure}[ht!]
\centering
\includegraphics[width=8cm]{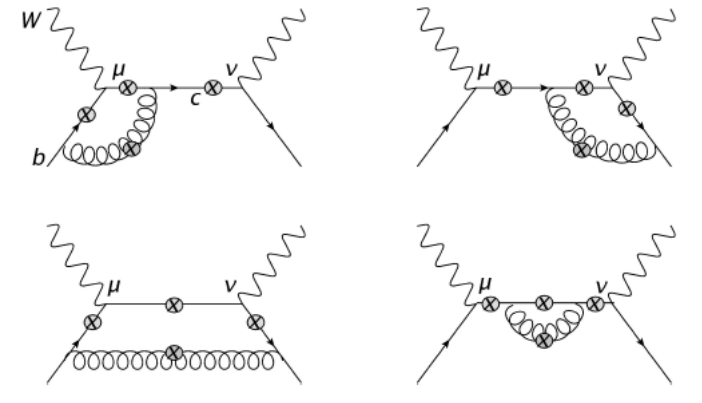}
\caption{Diagrams contributing to $\mathcal{O}(\alpha_s/m_b^2)$, (x) denotes a gluon insertion for symmetrization of HQET operators}
\label{fig:diags}
\end{figure}

The hadronic tensor can then be calculated from the imaginary parts of gluon-inserted diagrams of the forward-scattering amplitude as seen in Figure \ref{fig:diags}.

Additionally. the expansion around the on-shell b-quark in QCD must be matched onto HQET local operators. Analytic formulae are then obtained and unlike $\mu_{\pi}$, the operator for $\mu_G$ gets renormalized.

\subsection{Experimental Observables}
The relevant quantities used in the fit are the first 3 moments of the leptonic energy spectrum as a function of a cut on the lower energy limit,
\begin{equation}
\langle E_\ell^{n}\rangle_{E_\ell > E_{cut}} = \frac{\int_{E_{cut}}^{E_{max}} d E_\ell \ E_\ell^n \ \frac{d\Gamma}{d E_\ell}}
{\int_{E_{cut}}^{E_{max}} d E_\ell \ \frac{d\Gamma}{d E_\ell}}\ ,
\end{equation}
which are measured for $ n $ up to 4, as well as the ratio $R^*$ between the rate with and without a cut
\begin{equation}
R^* (E_{cut})= \label{eq:Rstar}
\frac{\int_{ E_{cut}}^{E_{max}} d E_\ell \ \frac{d\Gamma}{d E_\ell}}
{\int_0^{E_{max}} d E_\ell \ \frac{d\Gamma}{d E_\ell}}\ .
\end{equation}
 $R^*$ is needed to relate the actual measurement of the rate to one with a cut, from which one can then extract $|V_{cb}|$.
 
Due to the high degree of correlation in the first three linear moments, it is beneficial to instead study the central moments, including the variance and asymmetry of the lepton energy distribution. In our procedure we will consider only $R^*$ and the first 3 central moments,
\bea
\nonumber \ell_1(E_{cut})&=&\langle E_\ell \rangle_{E_\ell > E_{cut}}, \\ 
\ell_{2,3}(E_{cut})&=&\langle \left(E_\ell - \langle E_\ell \rangle \right)^{2,3} \rangle_{E_\ell > E_{cut}}\,.
\eea
In the case of the moments of the hadronic invariant mass distribution we similarly consider the central moments
\bea
\nonumber h_1(E_{cut})&=&\langle M_X^2\rangle_{E_\ell > E_{cut}}, \\
h_{2,3}(E_{cut})&=&\langle (M_X^2-\langle M_X^2\rangle)^{2,3}\rangle_{E_\ell > E_{cut}} .
\eea
  
\subsection{Results}
\begin{figure}[ht!]
\hspace{-0.8cm}
\\
\centering
\includegraphics[width=4.63cm]{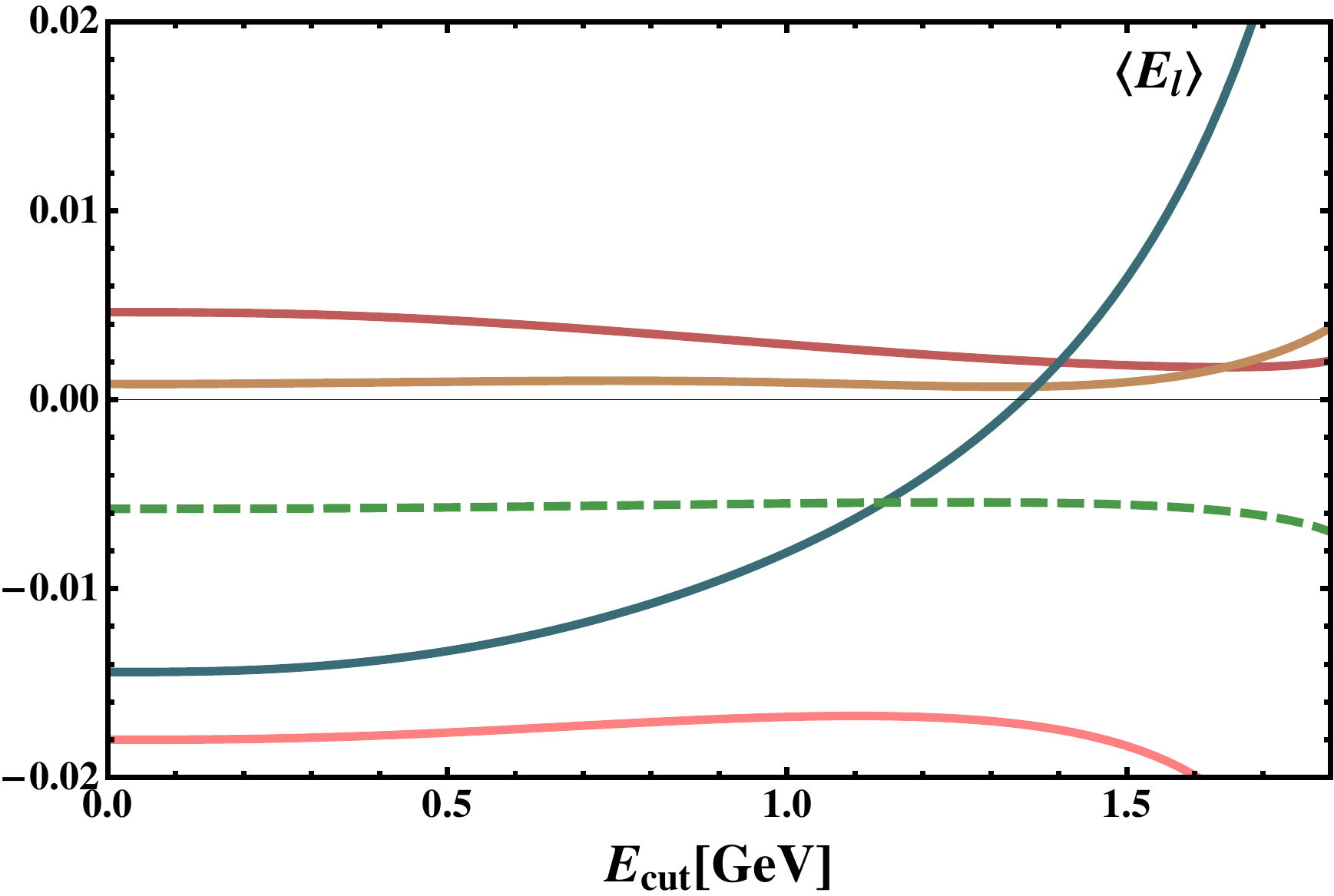}
\includegraphics[width=4.63cm]{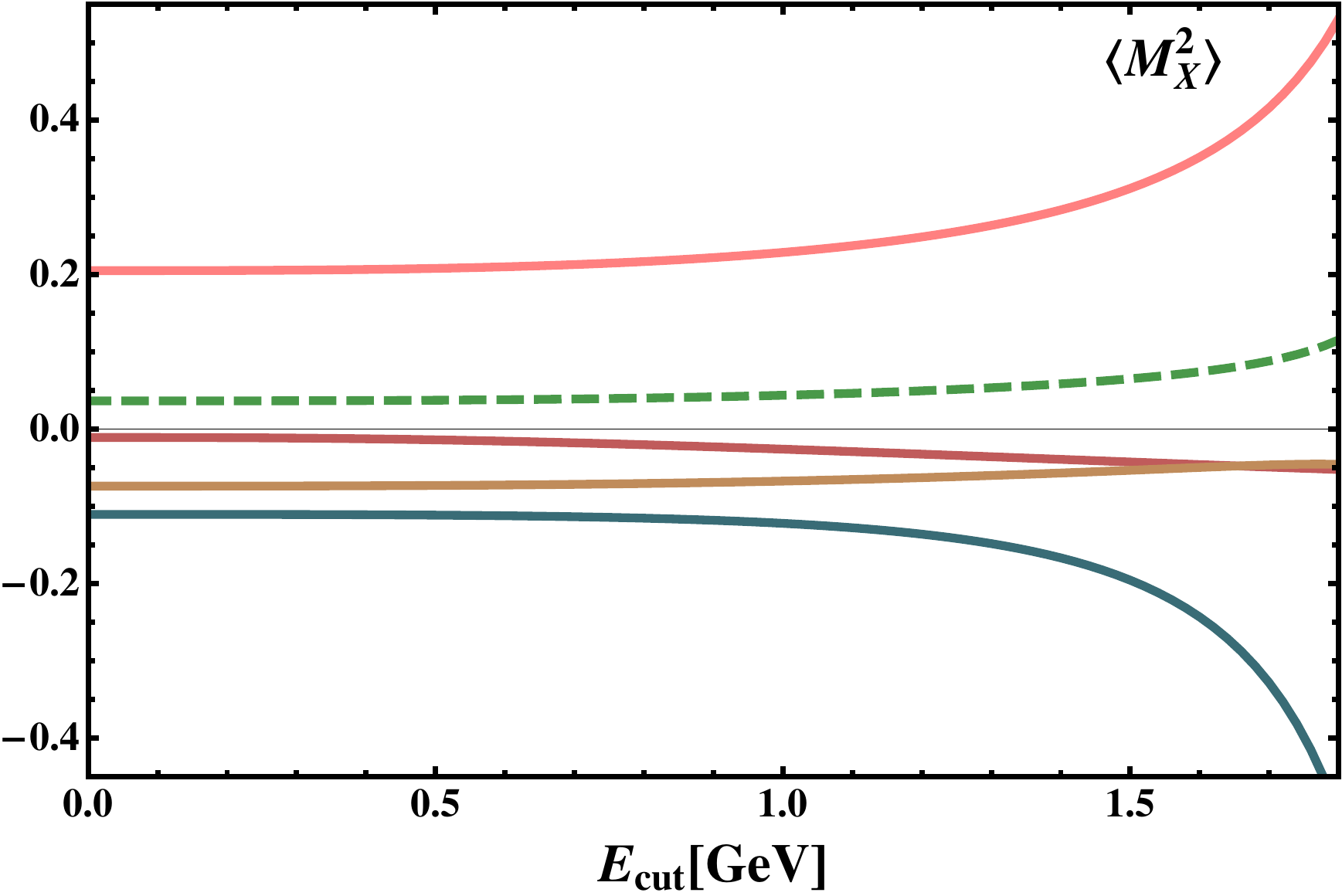}
\includegraphics[width=4.63cm]{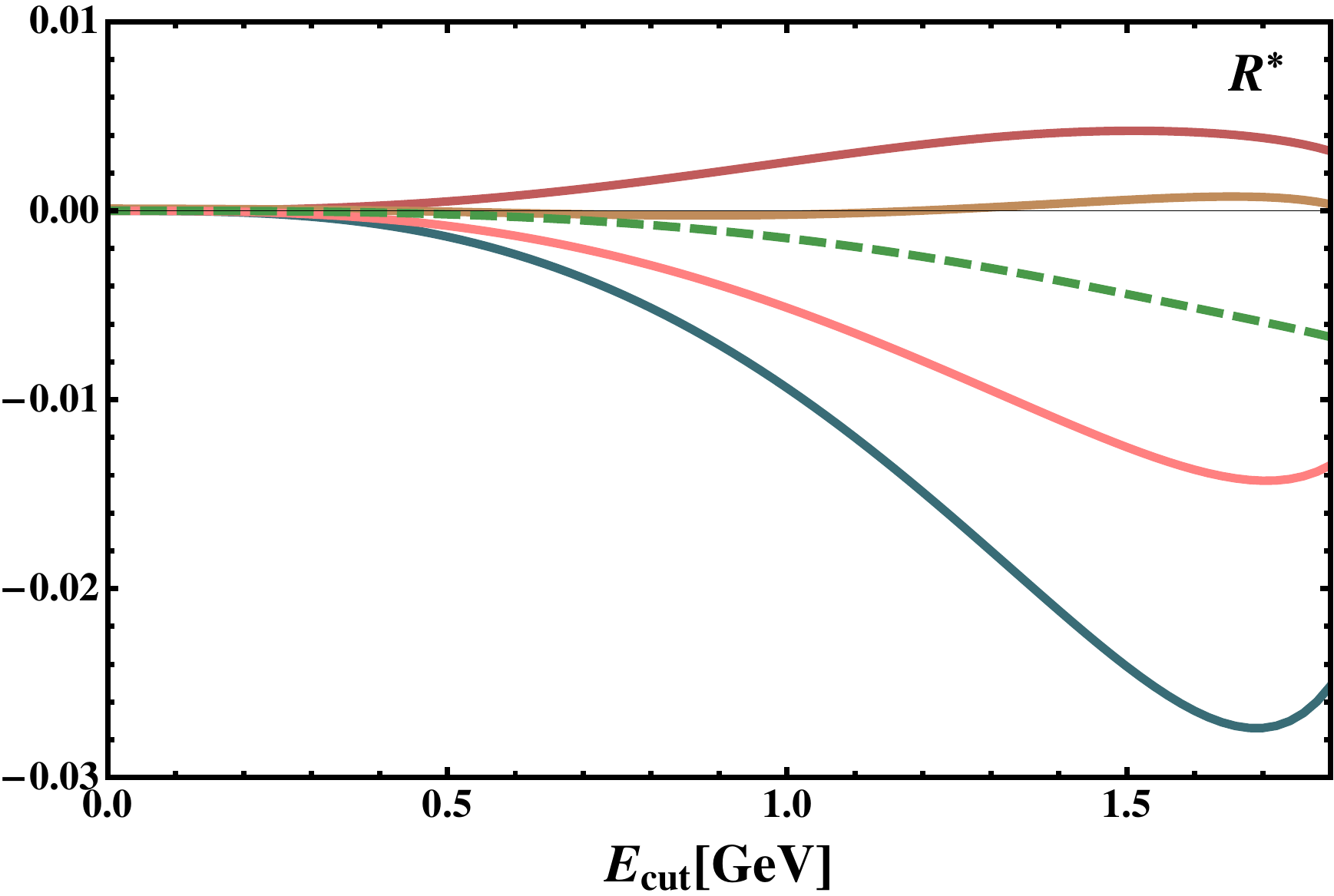}\hspace{0.0cm}\\
\includegraphics[width=4.63cm]{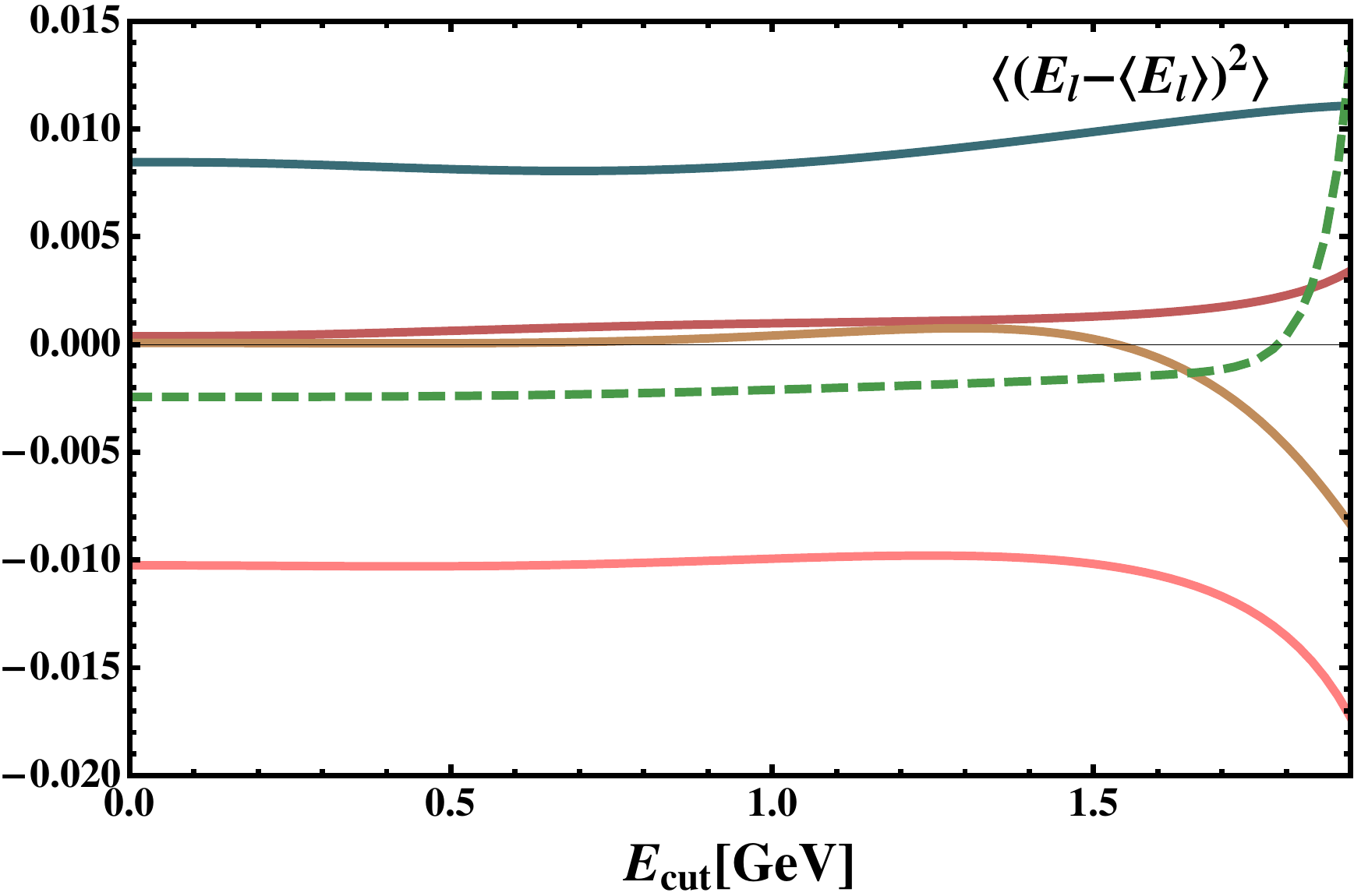}
\includegraphics[width=4.63cm]{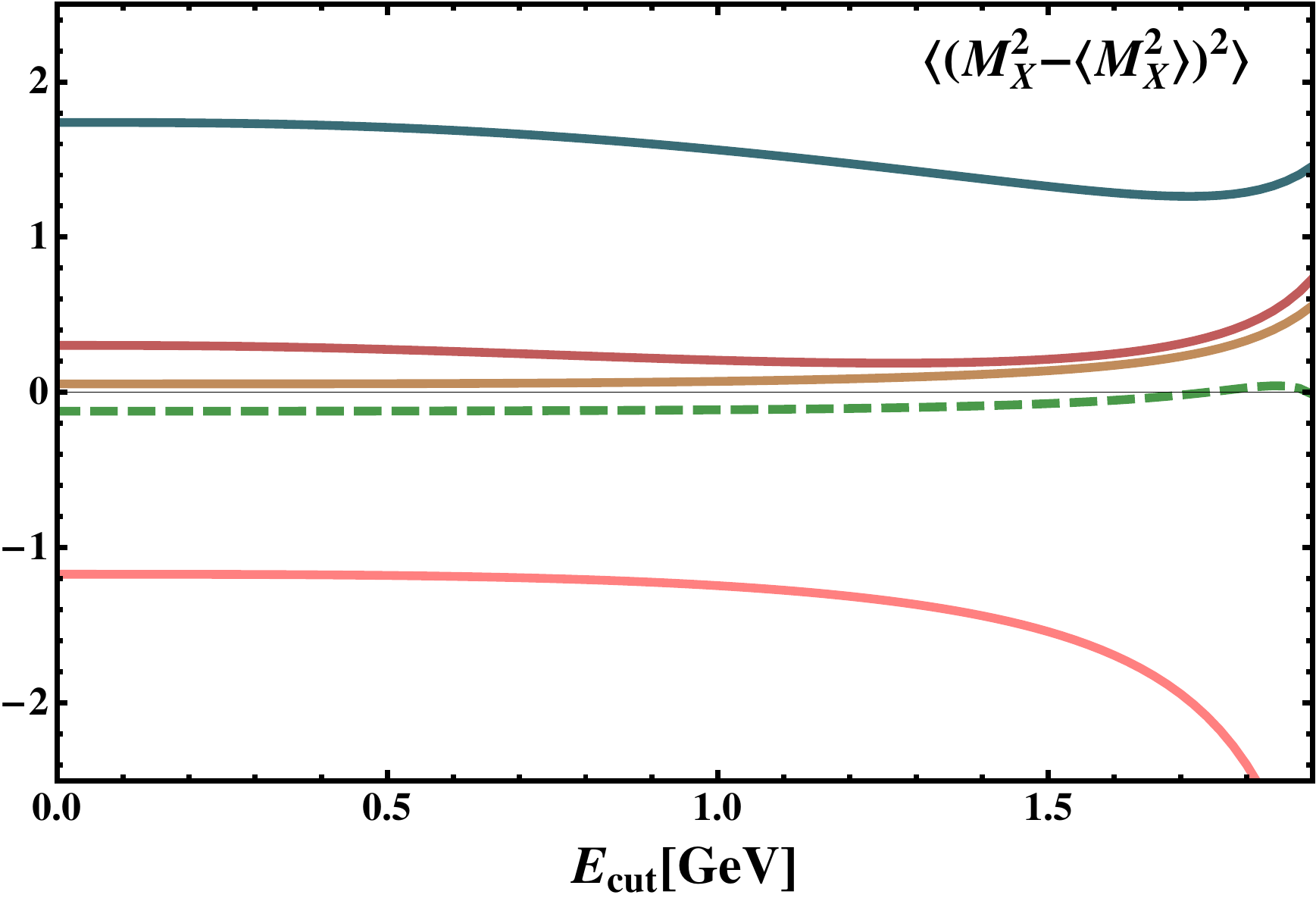}
\includegraphics[width=4.63cm]{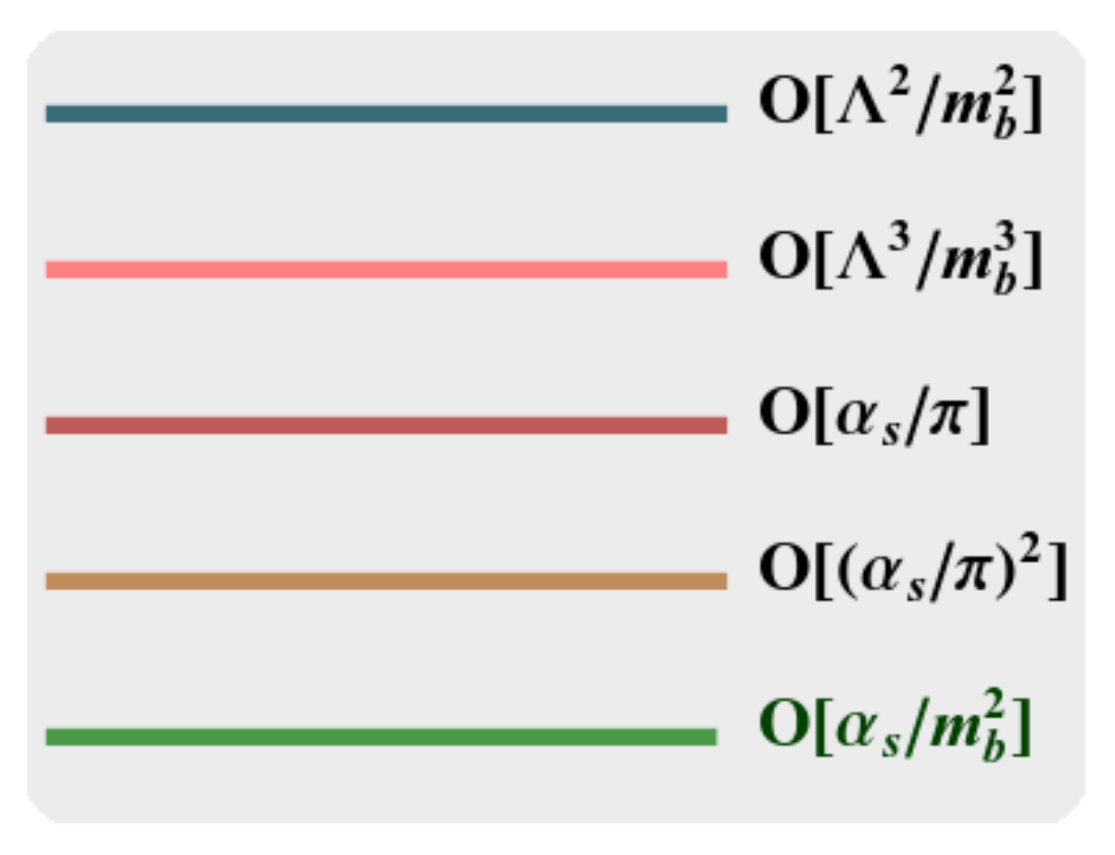}
 \caption{\sf \label{figgraphs:1} Contributions to some moments and observables from each correction. }
 \label{fig:1}
\end{figure}

We first present the results of the numerical integration for each of the theoretical observables as a function of the leptonic energy cut $E_c$ in Figure \ref{fig:1}. It is apparent that the new $O(\alpha_s \Lambda^2_{\rm QCD}/m_b^2)$ corrections are non-negligible, and compete with the NNLO contribution. This also drives the importance of continuing the calculation of the NLO corrections $O(\alpha_s \Lambda^3_{\rm QCD}/m_b^3)$ , as the leading order $1/m_b^3$ corrections have large coefficients, and so one would similarly expect the NLO correction to also be competitive. 

We perform the global fit following the same method as the previous determination, using for these proceedings only the case of \textbf{[D]} neglecting the external constraints on $m_b$ \cite{Alberti:2014yda}. The bottom row of Table \ref{results} is the results of the performed fit for the matrix element parameters, quark masses, and the resulting determination of \vcb. We find the new corrections lower \vcb by about 1\%, leading to $|V_{cb}| = (42.21 \pm 0.78) \times 10^{-3}$ \cite{Alberti:2014yda}.

\begin{figure}[ht!]
\centering
\includegraphics[width=8cm]{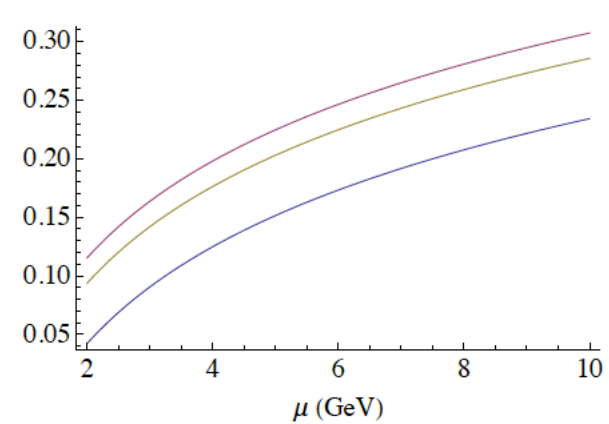}
\caption{Relative NLO correction to the coefficients of $\mu_G(\mu)$ in the width (blue), first (red) and second central (yellow) leptonic moments as a function of the renormalization scale $\mu$.}
\label{scaledep}
\end{figure}

\begin{tiny}
\begin{flushleft}
\begin{table}[ht!]
\begin{center} \begin{tabular}{|c|ccccccc|c|}
    \hline 
  &$m_b^{kin}$ & $ m_c$   &  $\mupi $ &$\rd$ &$\mug$ & $\rls$  & ${\rm BR}_{c\ell\nu}${\scriptsize (\%)}& $10^3 \,|V_{cb}|$ \\ \hline\hline
    $\mathcal{O}(\alpha_s^2, m_b^{-2}) $ & 4.541 &0.987 & 0.414 & 0.154 & 0.340 & -0.147 & 10.65 & 42.42 \\  
   {\scriptsize $\overline{m}_c(3{\rm GeV})$} & 0.023 & 0.013 & 0.078 & 0.045 & 0.066 & 0.098 & \ 0.16 & \ 0.86
   \\   \hline 
 &$m_b^{kin}$ & $ m_c$   &  $\mupi $ &$\rd$ &$\mug$ & $\rls$  & ${\rm BR}_{c\ell\nu}${\scriptsize (\%)}& $10^3 \,|V_{cb}|$ \\ \hline\hline
    $\mathcal{O}(\alpha_s^2, \alpha_s m_b^{-2}) $  & 4.553 &0.987 & \textbf{0.465} & 0.170 & \textbf{0.332} & -0.150 & 10.65 & 42.21 \\  
   {\scriptsize $\overline{m}_c(3{\rm GeV})$} & 0.020 & 0.013 & 0.068 & 0.038 & 0.062 & 0.096 & \ 0.16 & \ 0.78
   \\  \hline   
  \end{tabular}\end{center}
  \caption{A comparison between new(lower) and old(upper) global fits for $|V_{cb}|$ and kinematic parameters.}
  \label{results}
\end{table}
\end{flushleft}
\end{tiny}
Renormalization of $\mu_G(\mu)$ and $\mathcal{O}(\alpha_s \mu^2_G/m_b^2)$  lead to residual scale dependence seen in Figure \ref{scaledep}.  It is worth noticing that the lower the scale, the smaller the corrections.

\begin{figure}[ht!]
\centering
\includegraphics[width=10.0cm]{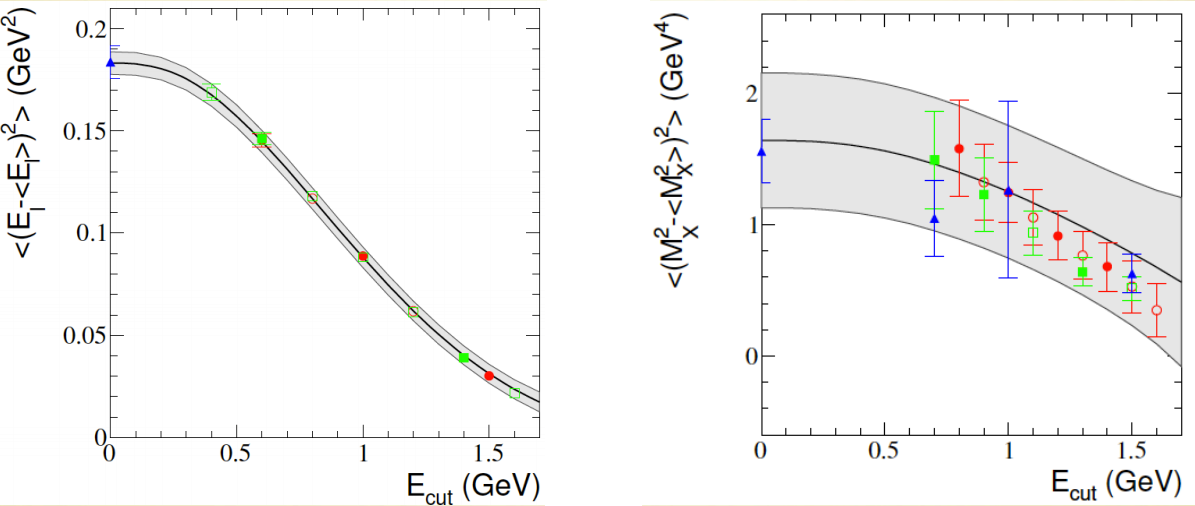}
\caption{Theoretical error bands with experimental points for varius observables.}
\label{theerr}
\end{figure}

Theoretical errors, as seen in Figure \ref{theerr}, are generally dominant in the fits. They are estimated in a conservative way by mimicking higher orders, varying the parameters by fixed amounts ($7\%$ for $\mu_{\pi,G}$, $30\%$ for $\rho_{D,LS}$). Quark-Hadron duality violation, which is expected to be suppressed, would appear as an inconsistency in the fit.
 
\section{The GGOU $|V_{ub}|$ Determination in Brief}
  We briefly discuss the current theoretical limitations on the inclusive $V_{ub}$ determination using the GGOU\cite{GGOU} method. We refer the reader to the relevant text for a more through understanding of the calculations involved.
  
  The hadronic structure functions in this case are dependent on unknown "shape-functions" $F_i$, that are convoluted with the perturbative hadronic functions,
\begin{equation}
 W_i(q_0,q^2) = m_b^{n_i}(\mu)
 \int dk_+ \ \textcolor{orange}{F_i(k_+,q^2,\mu)} \ W_i^{pert}
\left[ q_0 - \frac{k_+}{2} \left( 1 - \frac{q^2}{m_b M_B} \right), q^2,\mu \right].
\end{equation}

While these cannot be calculated from first principles, there exist a few ansatz that can be used to constrain the possible functional forms of $F_i(k_+)$. For example, there must be an exponential suppression as $k_+ \rightarrow -\infty$; it should be Positive Definite at tree level; there should be a hard cut-off, $\theta(\bar\Lambda-k_+) $; and that in principle the functions are non-universal, $(F_{1,2,3})$, when including any power corrections. Using these one can stipulate myriad functional forms that satisfy the constraints on the $k_{+}^n$ moments from the OPE. GGOU explores a variety of these, 

\begin{eqnarray}
\nonumber F_i(k_+) = N_i \,(\bar{\Lambda}-k_+)^{a_i} \, e^{b_i \,k_+}\
 \theta(\bar\Lambda-k_+)\quad\quad \quad\quad
{\mathtt (exponential)}\\
\qquad\qquad F_i(k_+) = N_i\,(\bar{\Lambda}-k_+)^{a_i}\, e^{-b_i \,(\bar\Lambda-k_+)^2} \, \theta(\bar\Lambda-k_+)
 \quad\quad{\mathtt (gaussian)}\\\nonumber 
\qquad\qquad F_i(k_+) = N_i e^{-a_i \left(\bar\Lambda -k_+ + \frac{b_i}{\bar\Lambda -k_+}\right)^2}\, \theta(\bar\Lambda-k_+)\quad\quad\quad\quad {\mathtt (roman)}.
\end{eqnarray} 

\begin{figure}[ht!]
\centering
\includegraphics[width=10.0cm]{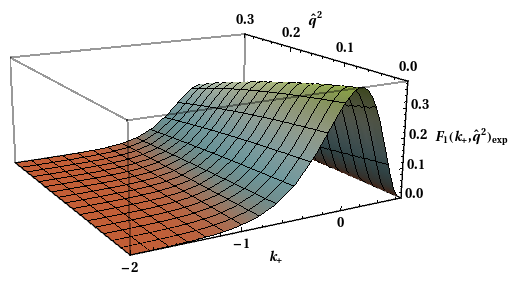}
\caption{Exponential Form for $F_1(q^2,k_+)$\textsl{•}}
\label{fig:f1}
\end{figure}

One can see an example solution using the exponential form for $F_1$ in Figure \ref{fig:f1}.
 
\begin{figure}[ht!]
\centering
\includegraphics[width=8.0cm]{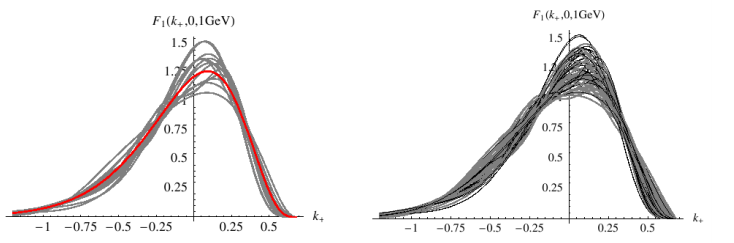}
\caption{Various 2-parameter and polynomial shape-function forms.}
\label{fig:formsw1}
\end{figure}

By varying these forms, including polynomial modifiers, we can gauge the contribution to the theoretical error from the uncertainty in our model choice. Figure \ref{fig:formsw1} shows the variation seen in these forms.
  
This method leaves room for improvement as it is model dependent and introduces bias. The theoretical error is again estimated by varying parameters.

\section{Conclusions}

For $V_{ub}$, the dominant theoretical errors are due to the high $q^2$ region. Goals to improving the inclusive determination with the GGOU method could come from removing the model dependence by using a statistically sound solution to the shape-functions (a neural-network solution is in development). The code also needs updating with the new fits for the kinematic parameters as well as the $\alpha/m_b^2$ corrections. Including constraints from the spectra from Belle-II would also help reduce the error.

\vcb will become more precise with the addition of the NLO calculations to the kinetic parameters at $\mathcal{O}(\frac{\alpha_s}{\Lambda}{m_b^3})$. Unfortunately, while we are confident in the applicability of this approach, steady increases in precision have not necessarily improved the $3 \sigma$ discrepancy between inclusive and exclusive determinations.

\end{document}